# The Shannon Entropy of a Histogram

S. J. Watts[*], L. Crow

Department of Physics and Astronomy, The University of Manchester, Oxford Road, M13 9PL, Manchester, UK

*Correspondence to: Stephen.Watts@manchester.ac.uk

**Abstract:** The histogram is a key method for visualizing data and estimating the underlying probability distribution. Incorrect conclusions about the data result from over or under-binning. A new method based on the Shannon entropy of the histogram uses a simple formula based on the differential entropy estimated from nearest-neighbour distances. Links are made between the new method and other algorithms such as Scott's formula, and cost and risk function methods. A parameter is found that predicts over and under-binning which can be estimated for any histogram. The new algorithm is shown to be robust by application to real data. A key conclusion is that the Shannon entropy of continuous data is $(1/2)\log_2 N$ bits which provides a limit to what one can learn from data.

## 1. Histogram Algorithm

The histogram is a key method for visualizing data and estimating the underlying probability distribution function (pdf). Incorrect conclusions about the data result from over-binning or under-binning. Computer software that automatically decides upon a suitable bin width can fool the unwary user. Scientists often adjust the bin size to get a visually appealing plot. The human visual system is very good at gauging when a histogram is under or over-binned. However, one must always be wary of a process that cannot be quantified. Two key parameters affect the choice of binning – the number of entries (N), and some variance measure of the pdf. Table 1.1 shows the different algorithms and the scientific software which uses them, refs. [1-13]. The final column indicates whether the algorithm uses a formula or cost function. Many of the algorithms minimise an Integrated Mean Square Error (IMSE) function, but arrive at different formulations using various approaches and approximations. The relationship between many of these algorithms and the one described in this paper will be discussed in Sections 4.2 and 5.0

The variables describing a histogram are as follows. A data point (or "entry") is assigned to a bin i where i = 1 to $N_{bin}$, which is the total number of bins. Defining $n_i$ as the number of entries in bin i then,

$$N = \sum_{i=1}^{i=N_{bin}} n_i \quad , \quad p_i = \frac{n_i}{N} \qquad (1)$$

N is the total number of points or entries and $p_i$ is the probability of counting events in bin i.

The Shannon entropy of the binned histogram, $H_B$, is

$$H_B = -\sum_{i=1}^{i=N_{bin}} p_i \log(p_i) \qquad (2)$$

The entropy is given in *bits* or *nats* when the log is base 2 or e respectively. This paper will use bits unless otherwise indicated. The entropy means that a minimal of $2^H$ bins are required to histogram the data. For example, a fair coin gives two discrete values – head or tail – each with a probability of 1/2. $H_B$ = 1 bits for this system which thus requires $2^1 = 2$ bins. A useful quantity is the efficiency, $\varepsilon$ which is defined as the ratio of the number of bins expected from the entropy, to the actual number of bins required to histogram all the data.

$$\varepsilon_H \equiv \frac{2^{H(bits)}}{N_{bin}} = \frac{e^{H(nats)}}{N_{bin}} \qquad (3)$$

For an unfair coin, $p_1 = 0.25$, $p_2 = 0.75$, $H_B = 0.811$ bits and $\varepsilon_H = 0.877$. As is well known, H is maximised when all the $p_i$ are the same, in which case, $\varepsilon_H = 1$. This is the uniform distribution.

Shannon entropy applies to discrete or categorical data. For continuous data, one must first assign the histogram variable, x, to the bin i. To do this, one needs to define the starting point of the histogram, $x_s$, which is set to be a sensible value close to the minimum value of x in the data. The bin index i for the data points $x_i$ with i = 1 to N is then,

$$i = \left\lfloor \frac{x_i - x_s}{\Delta} \right\rfloor + 1 \tag{4}$$

Where the floor function, $\lfloor\ \rfloor$, gives the greatest integer less than or equal to the real number between the brackets. $\Delta$ is the fixed bin width for all x. The choice of this bin width is crucial and is the main discussion of this paper. The differential entropy, h, is defined as,

$$h = -\int_S p(x) \log p(x) dx \tag{5}$$

where $p(x)$ is the probability density function and $x$ is the continuous random variable. S is the support of p(x). This is well-defined for a particular distribution. For example, the normal distribution with σ = 1 has h = 2.047 bits, Table 1.2 . The discrete entropy, H, and differential entropy, h, are related by [14],

$$H = h - \log(\Delta) \tag{6}$$

At first sight, for continuous data, H appears poorly defined because it is dependent on $\Delta$ which is not known. Incorrectly, one might think H to be infinite because $\Delta$ should tend to zero for a continuous distribution. $\Delta$ must be chosen so that the histogram properly represents the underlying distribution without bias. There is an extensive literature on the choice for $\Delta$ which is concisely summarised in reference [15]. A widely used formula for $\Delta$ is due to Scott, [2], and obtains the optimal bin width by minimizing the IMSE between the histogram estimate of the underlying distribution and the actual distribution. Scott's Formula is,

$$\Delta = \left\{ \frac{6}{\int_S p'(x)^2 dx} \right\}^{\frac{1}{3}} N^{-\frac{1}{3}} \tag{7}$$

where, $p'(x) = \frac{dp(x)}{dx}$. Note that this is infinite for a uniform distribution as $p'(x) = 0$. Eq. (7) cannot be used directly because one needs to know the pdf to apply it. This formula is often used on the assumption that the underlying distribution is normal, estimating the variance, $\sigma^2$, from the data. The resulting formula is called Scott's Rule.

$$\Delta = 2 \times 3^{\frac{1}{3}} \pi^{\frac{1}{6}} \sigma N^{-\frac{1}{3}} \tag{8}$$

Using Eqs. (5), (6), and (7) one can construct Table 1.2, which gives, h, $\Delta$ and H for various probability density functions. Table 1.2 shows that the entropy depends only on the logarithm of the number of events in the histogram. This is not a complete surprise. If all events were in separate bins, then the entropy would be maximal at $\log(N)$. However, for such a case, there would be no knowledge of the underlying distribution.

The values for entropy in the table also imply a small but non-zero value for $N = 1$ which may be traceable to the fact that Scott's Rule is an asymptotic result. Table 1.2 suggests that a histogram should have a well-defined entropy determined only by the number of entries, and will be of the form,

$$H_M = \frac{1}{M}\log(N) \tag{9}$$

with $1 \leq M \leq 3$. This assignment allows one to fix the bin width using Eq. (6) as,

$$\Delta = e^{h(nats)} N^{-\frac{1}{M}} = 2^{h(bits)} N^{-\frac{1}{M}} \tag{10}$$

This is can be applied to real data because one can estimate $h$ using the binless entropy estimator of Kozachenko and Leonenko, [16], for continuous distributions in Euclidean space. This is well described by Victor in [17]. For a one dimensional distribution,

$$h(bits) = \log_2(2(N-1)) + \frac{\gamma}{\log_e 2} + \frac{1}{N}\sum_{i=1}^{N}\log_2 \lambda_i \tag{11}$$

where, $\lambda_i$ is the nearest neighbour distance for the $i^{th}$ point in the histogram with value $x_i$. Eqs. (10) and (11) define the new algorithm for the bin width of the histogram.

To illustrate this algorithm, the Moyal distribution, see Appendix A1, was simulated for 10,000 entries with M = 1, 2, 2.6, 4 and 6. This is shown in Fig. 1 together with a fit for all values of M except 1. The Moyal distribution was chosen to illustrate the algorithm because it is a mix of a normal distribution with an exponential tail, and has no free parameters. The $\chi^2$/DF is 0.94, 1.31, 1.83 and 13.54 for M = 2, 2.6, 4 and 6 respectively. This example shows that M = 1 is over-binned as expected, that 2 < M < 3 produce a well binned histogram, and that once M > 3, the histogram is under-binned. These conclusions are for the data in Fig. 1 but apply in general as Section 4 will show.

2. **Entropy calculation using entries per bin**

A range of different distributions with a wide range of skewness and kurtosis were simulated to demonstrate the work in this paper; uniform distribution between zero and one, standard normal distribution, exponential with mean one, standard log-normal and an approximation to the Landau distribution due to Moyal. Appendix A1 provides more details. For brevity, these distributions will be referred to as uniform, normal, exponential, log-normal and Moyal in the rest of the paper.

**2.1 Entries per bin analysis**

An alternative way to calculate the entropy of the histogram is to count how many bins have 0, 1, 2, ...j entries per bin. One can then plot this distribution which will be called the "Entries per bin" or EntBin plot. This could itself be binned, but in the EntBin plots shown in the figures, each point will correspond to one value of j, which is why the phrase plot is used. Thus one can calculate the entropy in an alternative way.

$$H = \sum_{j=0}^{j=N_{Max}} N_j \times H_j = -\sum_{j=0}^{j=N_{Max}} N_j \times \frac{j}{N}\log\left(\frac{j}{N}\right) \tag{12}$$

Where, $N_j$ is the number of bins with j entries, and $H_j$ is the entropy associated with this entry, which is given above and uses the fact that the probability associated with j entries per bin is j/N. The maximum number of entries/bin is $N_{Max}$. The j = 0 bin has zero entropy, but this is included so that empty bins are not forgotten. The final unknown is the distribution for $N_j$. This can be written as,

$$N_j = N_S p(j;\mu) \tag{13}$$

Where $\mu$ is the average number of entries per bin and $p(j;\mu)$ is the probability for j entries per bin for the given $\mu$. The overall scale is determined by $N_S$ which will be discussed below. There are two possible values of $\mu$, either the expected mean given the entropy or the actual mean of the histogram, stated in Eq. (14) and Eq. (15) respectively.

$$\mu_H = \frac{N}{e^H} = \frac{N}{N^{\frac{1}{M}}} = N^{1-\frac{1}{M}} \qquad (14)$$

$$\mu_B = \frac{N}{N_{bin}} = \varepsilon \mu_H \qquad (15)$$

If the underlying pdf is a uniform distribution, then $N_S = N^{\frac{1}{M}}$, $\mu = \mu_H$ and $p(j;\mu) = \frac{1}{j!}\mu^j \exp(-\mu)$. This is a Poisson distribution. For large $\mu$ the distribution becomes normal with mean $\mu_H$ and standard deviation $\sqrt{\mu_H}$. The EntBin plot for simulated uniform, normal, and exponential distributions, for N = 10,000 are shown in Fig. 2a and Fig. 2b for M = 1.5 and M = 2.0 respectively. Some key observations from this figure.

- In Fig. 2a and M = 1.5, the uniform distribution gives an EntBin plot that is normal with expected mean and standard deviation, $10{,}000^{2/3}$ = 21.5 and sqrt(21.5) = 4.64, respectively. For M = 2, the uniform distribution gives an EntBin plot which is normal with expected mean and standard deviation, $10{,}000^{1/2}$ = 100 and sqrt(100) = 10, respectively.
- In Fig. 2a, for entries/bin > 9, the normal and exponential distributions give an EntBin plot with a normal distribution with mean $\mu_H$ but much larger standard deviation than expected. The underlying pdf's are smearing out the EntBin normal distribution, but the mean stays at $\mu_H$.
- In Fig. 2b and M = 2, for entries/bin > 9, the normal and exponential distributions give an EntBin plot which is smeared out, and is essentially uniform.
- Referring to the entries/bin variable as capital X. At low entries/bin (X +1 < 10), there is an excess of counts and this region of the plot can be described by a power law, $(X+1)^{-\alpha}$, with $\alpha$ in the range 1.4 to 2.0. A power law suggests scaling behaviour [18]. Since the number of entries/bin will scale inversely with bin size one would expect $\alpha$ to depend on the underlying pdf but not be too sensitive to M. Since the entropy associated with low values of X is small compared to the total, the contribution of this part of the plot to the total entropy will be discounted at this stage.

The above observations lead to a simplified mathematical model for the EntBin distribution. For a uniform pdf for large N, the EntBin plot will be described by a normal distribution. For all other pdf's, the EntBin plot will be uniform. Table 2.1 gives details of the fit to the data in Fig. 2, based on the observations above. One can approximate the summation in Eq. (12) with the integral,

$$H \approx N^{\frac{1}{M}} \int_0^{N_{Max}} \frac{X}{N} \log\left(\frac{X}{N}\right) p(X;\mu) dX \quad \text{nats} \qquad (16)$$

Note that p(X;μ) is now a pdf not a probability. There are two cases to consider.

**Case 1 - For an underlying uniform pdf**

p(X;μ) will be a normal distribution and the log(X) term will be almost constant around the mean and can be taken outside the integral. One can approximate Eq. (16) with Eq. (17).

$$H \approx \frac{(\log(\mu_H) - \log(N))}{\mu_H} \int_0^\infty X p(X;\mu) dX \qquad (17)$$

From which one immediately obtains,

$$H \approx (\log(\mu_H) - \log(N)) = \frac{1}{M} \log N \qquad (18)$$

This supports the assumption made in Eq. (9).

**Case 2 - All non-uniform pdfs or pdfs with a "shape"**

The previous model points to the importance that the mean entries/bin should be $\mu_H$ if one is to ensure that the entropy of the histogram will be as in Eq. (9). An EntBin plot with a uniform distribution with mean $\mu_H$ and stretched around this mean by a factor $s \times \mu_H$ is required. This also has a range, $R_X$, of $s \times \mu_H$. Thus Eq. (16) becomes,

$$H \approx \frac{1}{\mu_H} \int_0^{R_X} X \left( \log(X) - \log(N) \right) \frac{1}{R_X} dX \tag{19}$$

After some simple algebra and discounting small terms in $\frac{1}{R_X^2}$, one obtains

$$H \approx \frac{s}{2} \left( \frac{1}{M} \log(N) - \log(s) + \frac{1}{2} \right) \tag{20}$$

For consistency, Eq. (20) must lead to Eq. (9), so s = 2. The extra term is just -0.19 nats. The entropy in the low X entries due to the power-law contribution will balance out this term. This gives a key result that the maximum number of entries per bin will be approximately,

$$N_{Max} \approx 2\mu_H = 2N^{1-\frac{1}{M}} \tag{21}$$

Fig. 3a shows that this relation works reasonably well for many different distributions. This is the only value possible to ensure that the average entries/bin will be $\mu_H$ for those entries in the EntBin plot that contribute significantly to the overall entropy. These correspond to X > 10 in the EntBin plot. A key result can be obtained by re-arranging Eq. (21). Note the use of base 2 in the logarithm.

$$H(bits) \approx \log_2 \left( \frac{N}{N_{Max}} \right) + 1 \equiv H_X \tag{22}$$

For a uniform distribution the extra one bit is removed. Thus, the fact that a pdf has "shape" means one gains an extra bit. Eq. (22) defines $H_X$. One can estimate the M of any histogram in two ways; by calculating $H_B$ for the histogram using Eq. (2) and using this to estimate M, or calculate $H_X$ and then estimate M. These two estimates will be called $M_B$ and $M_X$ respectively. Specifically,

$$H_B = -\sum_{i=1}^{i=N_{bin}} p_i \log(p_i) \quad \text{and} \quad M_B = \frac{\log N}{H_B} \tag{23}$$

$$M_X = \frac{\log_2 N}{H_X} \tag{24}$$

Fig. 3b shows $M_X$ versus the input M for a normal, exponential, log-normal and Moyal distributions. These are very different distributions. For N = 10,000, there is good agreement between $M_X$ and the input M thus confirming the analysis in this section. For lower N, the $M_X$ value for the Moyal distribution at N = 100 and 1000, saturates at $M_X$ ~ 3 and 4 respectively. This is because there are not enough entries to achieve high values of $N_{Max}$. However, $M_X$ is reliable in the range of M that matters. For the data in Fig. 1, $M_X$ is 1.2, 2.0, 2.6, 2.9 and 5.5 for histograms in which M was set to 1, 2, 2.6, 3 and 6 respectively. The $M_X$ estimate is a quick way to estimate M from published histograms and will find application in Section 6.

Another way to check the assumptions underlying Case 2, is to estimate the mean number of entries per bin weighted by the entropy. This is defined as,

$$\mu_H^* \equiv \frac{\sum_{j=0}^{j=N_{Max}} j \times N_j H_j}{\sum_{j=0}^{j=N_{Max}} N_j H_j} \qquad (25)$$

Simulated data for the uniform, normal, exponential, log-normal, and Moyal distributions for N = 10,000 and M = 2 give $\mu_H^*$ as 100.8, 110.8, 118.9, 130.4 and 114.0 respectively. Even for the high kurtosis log-normal distribution, this mean is reasonably close to the expected 100.0.

To complete the discussion, one has to understand why $\sum_{j=0}^{j=N_{Max}} N_j = N_{Bin}$ but $N_S = N^{\frac{1}{M}}$ is used in the analysis above. This is easy to understand for Case 1 but is more subtle in Case 2. A complete description of the EntBin distribution for Case 2, with $0 \leq X \leq 2\mu_H$ is,

$$Counts(X) = \left(\frac{1-\varepsilon}{\varepsilon}\right) N^{\frac{1}{M}} \times \frac{1}{K_{Norm}} (X+1)^{-\alpha} + N^{\frac{1}{M}} \times \frac{1}{2\mu_H} \qquad (26)$$

The terms after the multiplication sign, ×, are the pdf's for the power-law at low X and uniform distribution that describe the two contributions to the EntBin plot. The discussion so far has only applied the second term in Eq. (26) because this contributes most of the entropy. The power-law function has to be normalized, which is given by, [18],

$$K_{Norm} = \sum_{j=0}^{j=20} (X+1)^{-\alpha} \qquad (27)$$

Integrating Eq. (26) gives the correct number of total bins,

$$N_{Bin} = \left(\frac{1-\varepsilon}{\varepsilon}\right) N^{\frac{1}{M}} + N^{\frac{1}{M}} = \frac{1}{\varepsilon} N^{\frac{1}{M}} \qquad (28)$$

In fact, Eq. (26) was written to ensure that it gave the correct total number of bins. Thus the "inefficient" bins are in the power-law part of the plot. From Eq. (26) one gets the number of empty bins (X = 0).

$$N_{Empty} = \left(\frac{1-\varepsilon}{\varepsilon}\right) N^{\frac{1}{M}} \times \frac{1}{K_{Norm}} \qquad (29)$$

This is a strong function of the efficiency. For example, for N = 10,000, the efficiency term in Eq. (29) changes from 1 to 13 for the normal and log-normal distributions respectively. This is due to the large kurtosis of the log-normal distribution. Inefficiency is inevitable with fixed width binning except for a uniform pdf.

Table 2.1 shows how well the EntBin plot in Fig. 2 is explained by the model described above by fits to the simulated data for uniform, normal and exponential distributions. Table 2.2 shows how well Eq. (29) agrees with data in Fig. 2. Eq. (29) is especially important as it links the number of empty bins with the efficiency, M and slope of the power law. A relationship that is not obvious without an understanding of the model that explains this plot.

## 3   Efficiency, $\varepsilon$, for different distributions

The efficiency of the histogram, $\varepsilon_H$ can be estimated using Eq. (3). It can also be estimated using h and the range of the data points, $R_{Data,}$ and it is simple to show that,

$$\varepsilon_H \equiv \frac{e^{H(nats)}}{N_{bin}} = \frac{e^{h(nats)}}{R_{Data}} \equiv \varepsilon_h \qquad (30)$$

Eq. (30) shows that efficiency is a characteristic of the distribution, but it does depend on N, due to the range. Both estimates of the efficiency can be obtained from the data, although $\varepsilon_H$ is more reliable as the range has a large error at small N. For pdf's with a known functional form – e.g. Table 1.2 – $R_{Data}$ can be estimated using the statistics of extremes, specifically, the mean range using Eq. (2.3) from ref. [19]. Figure 4 shows how the efficiency changes with N for uniform, exponential and normal distributions. These curves apply to any uniform, normal or exponential distribution as the efficiency is scale independent. The solid lines estimate $\varepsilon_h$, based on the pdf's known h and $R_{Data}$ calculated using the formula from ref. [19]. One standard deviation bands are drawn for each distribution. These were obtained from simulated data using standard deviation estimates from multiple trials. The error band for the uniform distribution can be estimated directly, because the range is well defined and thus the error on the efficiency is the same as the error on h. The variance on h is [20],

$$Var(h(p)) = \frac{1}{N} Var(\ln p(x)) + \psi_1(k) \quad \text{nats}^2 \quad (31)$$

Where k is the $k^{th}$ nearest neighbour used to estimate h, and $\psi_1$ is the trigamma function. The first term is zero for a uniform distribution, and $\psi_1(1) = \pi^2/6$. For k = 1, the error on h is $\delta h \approx 1.28/\sqrt{N}$ nats. This is a useful rule of thumb for other distributions and reduces for higher values of k. An estimate from multiple trials for the uniform distribution, $\delta\varepsilon \approx 1.5/\sqrt{N}$, is in good agreement with Eq. (31).

4. Discussion on the value of M

Section 1 showed that M must be greater than one. Its upper limit is fixed by Scott's formula which indicates a value of 3. This section investigates the optimal value of M and how it links to the methods of Scott, Rudemo, Stone, Shimazaki and Shinomoto.

4.1 M and its relation to over and under-binning

For a specific set of data the value of h is calculated using Eq. (11). Once M is chosen, the bin width $\Delta$ is calculated using Eq. (10). One can then calculate the actual histogram entropy. One thus has two estimates of the entropy. $H_M$ given by Eq. (9) which is the input entropy, and $H_B$ given by Eq. (2) which is the actual entropy of the histogram. For consistency these should be the same. To test how well this works, one takes the ratio,

$$R = \frac{H_B}{H_M} \quad (32)$$

There is a closed form equation for the case when the data is uniformly distributed which is also a good approximation for other pdfs. Using the results at the start of Section 2.1, if the underlying pdf is a uniform distribution, then $N_S = N^{\frac{1}{M}}$, $\mu = \mu_H$ and $p(j;\mu) = \frac{1}{j!}\mu^j \exp(-\mu)$ in Eq. (12) one can show that

$$R = \frac{H_B}{H_M} = \frac{M}{\log(N)} \sum_{j=1}^{j=N_{Max}} \frac{j}{j!} \log(\tfrac{N}{j}) N^{(1-\frac{1}{M})(j-1)} \exp(-N^{(1-\frac{1}{M})}) \quad (33)$$

Poisson fluctuations cause over-binning and empty bins and thus reduce $R_H$ below one between 1 < M < 2. This is illustrated in Fig. 5. Fig. 5 shows the value of R for uniform and normal distributions for values of N = 50, 500 and 10,000. R from Eq. (33) is also shown in Fig. 5a and 5b. R plateau's close to one for $M \geq 2$, once N > 50. Eq. (33) is exact for Fig. 5a and works well for the normal distribution in Fig. 5b. This is because $\mu_H$ is the correct mean to use for any distribution although some smearing of the assumed Poisson distribution will occur. Fig. 6 is a repeat of Fig. 5 except that the efficiency, $\varepsilon_H$, Eq. (3) is plotted rather than R. The N = 500 simulation is not shown, to aid the clarity of this figure. The binned efficiencies are consistent with the $\varepsilon_h$ estimates in Fig. 4. There are two key features in Figs. 5 and 6.

For N > 50, both R and ε saturate beyond M > 2. R is one for large N and close to one even for N = 50. For M < 2, Poisson fluctuations affect the histogram. Over-binning sets in for M < 2.

R and especially $\varepsilon_H$ show variation at larger values of M. This is more evident for low N. $\varepsilon_H$ is more sensitive due to the $e^H$ factor. This is the onset of under-binning and is dependent on N. For pdf's with "shape", this can be understood by consideration of Eq. (26). From Eq. (26), the average EntBin counts for an entry $N_j$ will be,

$$\mu_c = N^{\frac{1}{M}} \times \frac{1}{2\mu_H} = \frac{1}{2} N^{\left(\frac{2}{M}-1\right)} \tag{34}$$

Due to Poisson statistics the fraction of $N_j$ entries greater than zero will be,

$$F_{Entries} = 1 - \exp(-\mu_c) \tag{35}$$

M = 2 is a unique value since $\mu_c$ = 0.5 independent of N and $F_{Entries}$ ~ 40%. By M = 3, $F_{Entries}$ has dropped significantly to 13%, 6% and 2% for N = 50, 500 and 10,000 respectively. Although there are more $N_j$ values at higher N, the estimate of the entropy from Eq. (12) is dependent on a smaller fraction of these terms for M > 2. At higher M, the number of bins also gets smaller. This is illustrated in Fig. 6b. At M ~ 2.5, for N = 50, there are 7 bins and $\varepsilon_H$ jumps upwards at the transition between 8 and 7 bins. As M increases, the entries redistribute themselves between these 7 bins, and $\varepsilon_H$ decreases because the estimate is affected by the reduced $F_{Entries}$, until it reaches a minimum. It then jumps up again when the number of bins changes to 6. This repeats, with the size of the fluctuations increasing as M increases because $F_{Entries}$ continues to fall. The same effect can be seen for the N = 10,000 simulation, but this happens at higher M. This behaviour will always be seen for M > 3 because as will be shown later, the cost function increases beyond this point, which corresponds to under-binning.

Fig. 6a shows that these fluctuations also appear for the uniform distribution for M > 2 even for N = 10,000. Since the uniform distribution is flat, the optimal point for M is when the Poisson fluctuations are gone, which occurs at M = 2.

The overall conclusion is that M = 2 is the minimum required to avoid over-binning (Poisson fluctuations) and that M should be in the range 2< M < 3.

**4.2 M and its relation to cost and risk function algorithms**

Many histogram algorithms minimize a cost or risk function to choose the bin width. We will start with the method due to Shimazaki and Shinomoto, [9]. Their cost function penalises over-binning and under-binning. The underlying method is the use of IMSE. Their cost function is,

$$C_{SS} = \frac{2\mu_B - \sigma_B^2}{\Delta^2} \tag{36}$$

$\mu_B$ and $\sigma_B$ are the mean number of events per bin and its standard deviation respectively. In ref. [9] the cost function is evaluated for different bin sizes. The bin size for the lowest cost function is chosen as the best value.

Fig. 7a, 7b, and 7c show this cost function and efficiency for uniform, normal and exponential distributions, for N = 500, respectively. The cost function and efficiency have been normalized. The cost function was normalized with the value at M = 2 shifted to zero and the range between M = 1 and M = 2, set to one. An offset of 0.05 has been added to allow a log scale to show any values that went negative. The efficiency was normalized by setting the M = 2 value to 1.0 The actual value of the efficiency is given in the caption.

Fig. 7 illustrates that as expected, the efficiency saturates at M = 2. The efficiency also indicates under-binning beyond M = 2.5 and M = 2 for the normal/exponential and uniform distributions respectively. The cost function drops dramatically from M = 1 to M = 2 with a minimum between 2 and 3 for the normal and exponential distributions. This rapid drop between M = 1 and M = 2 is caused by the reduction in Poisson fluctuations. For the uniform distribution, the cost function levels off beyond M = 2. Since the uniform distribution has no shape the cost function cannot increase once Poisson fluctuations are removed.

Again, a key conclusion is that for the uniform distribution, M = 2 is a unique choice, for N > 50. Fig. 7d repeats the exercise but with the Moyal distribution and N = 10,000, to compare with the actual histogram in Fig. 1. This again confirms the link between the value of M and whether the histogram is correctly binned. Note that the efficiency starts to fluctuate for M > 2.5 which indicates under-binning, which is also the point at which the cost function starts to increase from minimum. The curves are a fit using Eq. (38) which is described below.

In Fig. 8, the normalised cost function for a uniform, normal, exponential, log-normal and Moyal distributions for N = 10,000 are shown. A universal curve is seen between M = 1 and M = 2. Note again the lack of any increase in the cost function beyond M = 2 for the uniform distribution because it has no shape. The behaviour of the normalised cost function between M = 1 and M = 2 can be explained, using the underlying formulae that describe the entropy algorithm, with an equation that is independent of the underlying pdf. The formula is derived in Appendix, Section A2. It is,

$$C_{NR} \approx \frac{\varepsilon}{\varepsilon_1} N^{(\frac{1}{M}-1)} - \frac{\varepsilon_2}{\varepsilon_1} N^{-\frac{1}{2}} = \frac{m_1}{\mu_H} - m_2 N^{-\frac{1}{2}} \qquad (37)$$

Eq. (37) implies that one would expect $m_1$ and $m_2$ to be around one. The normalised cost goes to zero at M = 2. It will only increase due to the shape of the pdf being lost due to under-binning. This is seen in Figs. 7 and 8. The minimum is always between M = 2 and M = 3 with the M = 2 value being within a few percent of the actual minimum. Fig. 8 clearly shows the increase in the cost beyond M = 3 for all distributions apart from the uniform. Fluctuations in the cost are due to under-binning. This is also apparent for the uniform distribution. The explanation is similar to that for the fluctuations in the efficiency described in Section 4.1. To fit these curves, the following form was used based on Eq. (37) plus an empirical term for the increase in the cost beyond M = 3. Note the offset of 0.05 to allow zero values to be seen on a log-scale graph.

$$C_{NR} = m_1 N^{1/M - 1} - m_2 N^{-1/2} + m_3 (M - |m_4|)^2 + 0.05 \qquad (38)$$

The results of the fits are shown in Table 4.1. $m_1$ and $m_2$ are close to one for all distributions. The fit to the data is between M = 1 and M = 4 but is extrapolated to M = 6. This was due to the behaviour of the cost for the log-normal distribution. The cost function rise is similar for the normal, exponential, and Moyal distributions. The log-normal has a minimum at M = 2. This is because of the large kurtosis – a large peak to tail ratio. It also has a large rise in cost beyond M = 2.5. This type of distribution is discussed in more detail in Section 6.

The work of Shimazaki and Shinomoto is also useful in that it applies to the uniform distribution, which the Scott formulation does not. Ref. [9] shows there are two asymptotic solutions for the cost function, depending on the behaviour of the autocorrelation function of the pdf,

$$\varphi(\tau) = \int_S p(x) p(x - \tau) dx \qquad (39)$$

If the autocorrelation is symmetric about zero then the N dependence is of the form, M = 3. The formula for the bin size is,

$$\Delta \sim \left[ -\frac{6}{\phi''(0) N} \right]^{1/3} \qquad (40)$$

The differentiation of $\phi$ is with respect to $\tau$. After some algebra, Appendix A3, it can be shown that this equation is identical to Scott, ref. [2] and Eq. (7). Scott's work is extended because if the autocorrelation has a cusp at zero shift then the N dependence implies an M = 2 dependence. In this case, the formula for the bin size is,

$$\Delta \sim \left[ -\frac{3}{\phi'(0+)N} \right]^{1/2} \tag{41}$$

Thus the fact that the cost function is minimum at M = 2 for a uniform distribution, and between M = 2 and M = 3 for other distributions as shown earlier in this section, is consistent with the discussion of Shimazaki and Shinomoto on the asymptotic solutions to the cost function. The log-normal with its large kurtosis also prefers M = 2.

An alternative approach is to use the risk functions of Rudemo, [5], and Stone, [6]. They arrive at the same function by different routes. Their function can be written as,

$$R_{RS} = \frac{1}{\Delta}\left( \frac{2}{N} - \sum_{i=1}^{i=N_{Bin}} p_i^2 \right) \tag{42}$$

Simple algebra shows that the cost function and risk functions are equivalent. They are related by

$$R_{RS} = \frac{R_{Data}}{N^2} C_{SS} - \frac{1}{R_{Data}} \quad \text{where } R_{Data} \text{ is the range of the data points, which is } N_B \Delta. \tag{43}$$

The parameter M which controls the histogram entropy, links the work of Shimazaki and Shinomoto, Rudemo, Stone and Scott.

## 5. The method of Knuth and early algorithms

Knuth, ref. [21], employs a very different method to those described in Section 4. The maximum of a posterior probability function, $P_K$, is calculated for different numbers of bins and the maximum found. The function is,

$$\log(P_K) = N \log B + \log \Gamma\left(\frac{B}{2}\right) - B \log \Gamma\left(\frac{1}{2}\right) - \log \Gamma\left(N + \frac{B}{2}\right) + \sum_{k=1}^{k=B} \log \Gamma\left(n_k + \frac{1}{2}\right) \tag{44}$$

To make the formula easier to read, $N_{Bin}$ is replaced with B. $n_k$ are the number of entries in each bin k. To understand this in the context of the entropy approach, this function was calculated using the Matlab code given by Knuth, and at the same time the resulting histogram entropy was calculated so that M could be estimated using $M_B$ given in Eq. (23). Fig. 9 shows the Knuth function as a function of $M_b$ for a normal distribution with N = 500. The normalised cost function is also shown. Knuth's function mirrors the cost function and peaks between M = 2 and 3. Eq. (44) can be simplified by approximating $\log\Gamma(x)$ and after some algebra, the last term is found to be related to the histogram entropy, subject to an approximation requiring $n_k > 0$. The approximated function, Eq. (45), is also plotted. It works well for M > 1.5, which is acceptable as suitable binning would require a larger value of M. It fails below M = 1.5 due to the $n_k > 0$ criterion. The approximation is,

$$\log(P_K) \sim \left(N + \frac{B}{2} - \frac{1}{2}\right) \log\left(\frac{B}{N + \frac{B}{2}}\right) + \frac{1}{2}\log 2 + N(\log N - H_B) \tag{45}$$

It is interesting to note that the histogram entropy also enters this function. This method is easier to implement with Eq. (45) and it is useful to plot the function versus $M_B$ which has been shown to control the over and under-binning of the histogram.

Algorithms related to Sturges' rule are left to consider from Table 1.1. Ref. [2] is by Scott who explains the rule better than the original paper. Since this rule and variants have been superseded by significantly better algorithms, they are not recommended for automated binning in scientific software, which is also the view of Scott in ref. [2].

## 6. Applications

Four examples are given that illustrate some key issues that arise when a histogram is generated.

### 6.1 Re-coding the data to achieve higher efficiency

Fig. 8 illustrated that the log-normal pdf with its high kurtosis is difficult to estimate with a fixed bin width. For N = 10,000, the efficiency is just 7%. Fig. 10a shows the cost function and efficiency for N = 500. The efficiency is ~ 23%, which is higher due to the lower value of N. Note that the efficiency is not stable after M = 2. It continues to grow as the tail of the distribution is better reconstructed with a larger bin size but at higher cost. Shannon's coding theorem, [14], states there must be a code that efficiently uses the entropy of the data when transmitting it from source to receiver. For a histogram this means re-coding the data by transformation of the variable x. Two options are equiprobable binning or applying the Box-Cox transformation, [22], to map to a normal distribution. The re-coded data then has higher efficiency. Equiprobable binning is simple to implement. This distribution is uniform so that M should be set to 2. The data is ordered and then divided into $N^{1/2}$ bins. The Box-Cox transform for the log-normal distribution transforms x to log(x). For both options the bin edges are identified and then the pdf determined – probability per bin width. Fig. 10b shows the log-normal distribution for N = 500 for fixed bin size, Box-Cox transformed, and equiprobable binning. In all cases M = 2, so the entropy is the same. The cost and efficiency versus M would be as Fig. 7a and 7b for the equiprobable and Box-Cox re-coded data respectively. A two parameter fit was made for all three binning methods. Fixed binning has 96 points, only 51 are non-zero, and $\chi^2$/DF = 1.11. The normal distribution binning has 30 points, with a $\chi^2$/DF of 0.87. Finally, the equiprobable binning has 22 points with a $\chi^2$/DF of 1.25. Fig. 10b shows that the re-coded data provides a much improved description of the pdf with the efficiency increasing from 23% to 100% and 75% for equiprobable and Box-Cox transform respectively.

### 6.2 Typical data – example due to Wand

Wand, [23], improved Scott's formula, although the algorithm is not easy to code. Fig. 11a shows the income data, with N =7202, referenced and used by Wand, with the three bin sizes used in Fig. 3 of [23]. These correspond to M = 2.4 (Wand algorithm), M = 2.8 (Scott's Rule), and M = 5.6 (S-Plus default binning). This illustrates the danger of using software default binning if it uses a poor algorithm. Fig. 11b shows the normalised cost and efficiency versus M. Wand's binning is clearly sensible. Scott's Rule is slightly under-binned and the S-Plus choice is very under-binned. There are repeated values in this data, nevertheless, ~ 91% of the data are usable for the estimate of h. This is a common problem which is due to the resolution of the original data. In this case, this is not a serious problem in terms of estimating h. By identifying the maximum entries/bin one can estimate $M_x$ using Eq. (24). $M_x$ is 2.4, 2.75 and 5.1 for each bin choice, in good agreement with the actual values used. This method can be applied directly to the histograms in Fig 3 of ref. [23] – with the same results - and is a good method to quantify the quality of a histogram in any publication.

### 6.3 Old Faithful data

A well known dataset from ref. [24] with N = 272 has poor resolution data with a significant number of repeated values. Only a few percent can be used to determine h for the nearest neighbour estimate. The solution is to use the $k^{th}$ nearest neighbour (kNN) to get a robust estimate of h. The differential entropy estimate of Konozenko and Leonenko has been updated for arbitrary k in ref. [25]. Eq. (11) is replaced by,

$$h = \log(2(N-1)) - \psi(k) + \frac{1}{N}\sum_{i=1}^{i=N}\log(\lambda_i^k) \tag{46}$$

where $\lambda_i^k$ is the $k^{th}$ nearest neighbour distance to the $i^{th}$ point and $\psi(k)$ is the digamma function. Fig. 12a shows the estimate of h versus the kNN choice and also the fraction of points used. This illustrates the poor resolution of this data. However h is stable for kNN > 8 once more than 60% of the data is used. Fig. 12b shows the cost and efficiency for this data. The repeated values make these variables meaningless for M < 1.5. The plot does show that M = 2 is a sensible choice for the histogram. Fig. 12c is the histogram for M = 2 and shows the two peak structure of the data. The $M_x$ value for this histogram is 2.09. Most algorithms would have difficulty with this data. The combination of Eq. (46) and a safe choice of M = 2, shows the robustness of the new algorithm.

### 6.4 Galaxy red-shift data collated by Wasserman

The previous example has a mix of two distributions. This is not unusual in data. A good example is taken from ref. [26] due to Wasserman. The galaxy red-shift data (N =1266 for Z < 0.2) consists of multiple narrow peaks. There are a small fraction of repeated values (10%) so the estimate of h uses Eq. (46). Fig. 13a shows h and also the fraction of points used for the estimate, versus kNN. Fig. 13b shows Rudemo and Stone's risk function and also the risk predicted by the cost using Eq. (43) versus M to show these two formulations give identical results. The risk function is shown because this method is used by Wasserman. The estimate of h is based on the log of the nearest neighbour distance, and thus it is good at identifying fine structure in data which other measures of spread could miss. The minimum of the cost function is at M = 1.5 due to the fine structure, but this then suffers from Poisson fluctuations. A value of M = 2 is best because the fine structure is seen but the histogram does not suffer from Poisson noise. Fig. 13b also shows the efficiency versus M which is not stable due to the multiple peak structure of the data. Fig. 13c shows the histogram for M = 1.5 (241 bins), M = 2.0 (74 bins) and M = 4.5 (10 bins) as used in [26]. Wasserman chooses 73 bins, which is not actually the minimum point of the risk function, however, it is the better choice, which a knowledge of M informs. The $M_x$ values are 1.86, 2.53 and 3.51. Despite the structure of the data, these values are still indicative of the quality of the binning.

### Conclusions

The Shannon entropy of a histogram can be written as $(1/M)\log N$, where N is the number of entries and $M \geq 1$. As a consequence one can calculate the bin width, Eq. (10), which is evaluated by estimating the differential entropy of the data using a method originally proposed by Kozachenko and Leonenko. M is shown to control whether a histogram is over or under-binned for any probability density function. If M < 2 then the histogram is over-binned and subject to Poisson fluctuations. If M > 3 then the histogram is under-binned. It is recommended to start with the largest entropy before Poisson fluctuations arise, and thus set M = 2. M can then be increased to a value between 2 and 2.5 using the cost function, Eq. (36), and histogram efficiency, Eq. (3) as a guide. M can be estimated for any histogram, whatever the algorithm employed to determine the bin width, by finding the maximum number of entries per bin, and then evaluating $M_x$ using Eq. (24). Alternatively, the entropy of the histogram can be calculated and $M_B$ determined using Eq. (23). $M_X$ uses just two parameters of the histogram, maximum entries per bin and total entries, to give a "Goldilocks" type statistic which tells whether the histogram is over (M < 2) or under-binned (M > 3), or just right ($2 \leq M \leq 3$). No such statistic exists at present and is a useful guide to any scientist when generating a histogram.

M should always be set to 2 for a uniform distribution. Using Eq. (10), the bin width is just the range divided by $\sqrt{N}$. This is the square root algorithm used by Microsoft's Excel spreadsheet package, [10], for which no published explanation has been found. Shannon's coding theorem shows there is a code that efficiently uses the entropy of the data. For a histogram, this means re-coding the data by either equiprobable binning, or mapping to a normal distribution by applying the Box-Cox transform. A fixed bin width histogram with low efficiency can then be transformed to achieve an efficiency of 100% (uniform) or greater than 50% for N < 10,000 (Box-Cox, normal). This will lead to an improved description of the pdf.

The new algorithm is an important building block for estimation of information measures from data. Ref. [27] uses the new algorithm to fix all continuous variables to the same entropy, setting M = 2. This is important to avoid bias when estimating the mutual information from binned data.

The probability density function is an idealized concept never achieved for real data since this requires an infinite number of samples. For one variable, the Shannon entropy is limited to (1/2)log$_2$N bits. Data is finite and this limits what one can learn. For example, because the mutual information between two variables cannot be more than the entropy of one of the variables, it cannot be larger than (1/2)log$_2$N bits for continuous data. This limitation is not due to the estimation method but is a fundamental limit. Ref. [28] derive an O(log(N)) limit based on other arguments.

**Acknowledgements**

LC thanks UKRI (STFC) and The University of Manchester for research student funding. SW thanks the Leverhulme Trust for their support with an Emeritus Fellowship.

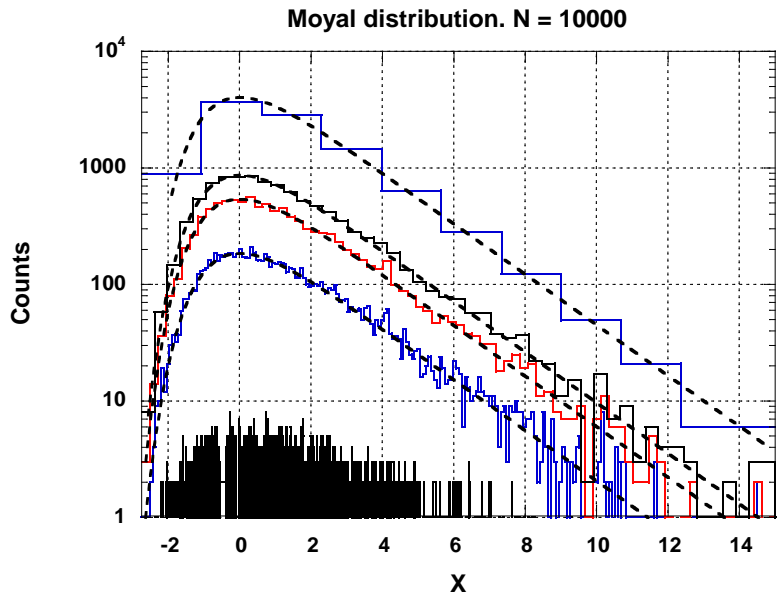

Fig.1 Entropy histogram algorithm applied to simulated data for a Moyal distribution with N = 10,000. For increasing peak value, M = 1, 2, 2.6, 3 and 6. A fit using the Moyal distribution, Appendix A1, was made to the histograms, except M = 1, giving $\chi^2/DF$ of 0.94, 1.31, 1.83 and 13.54. Note that the Moyal distribution has no free parameters.

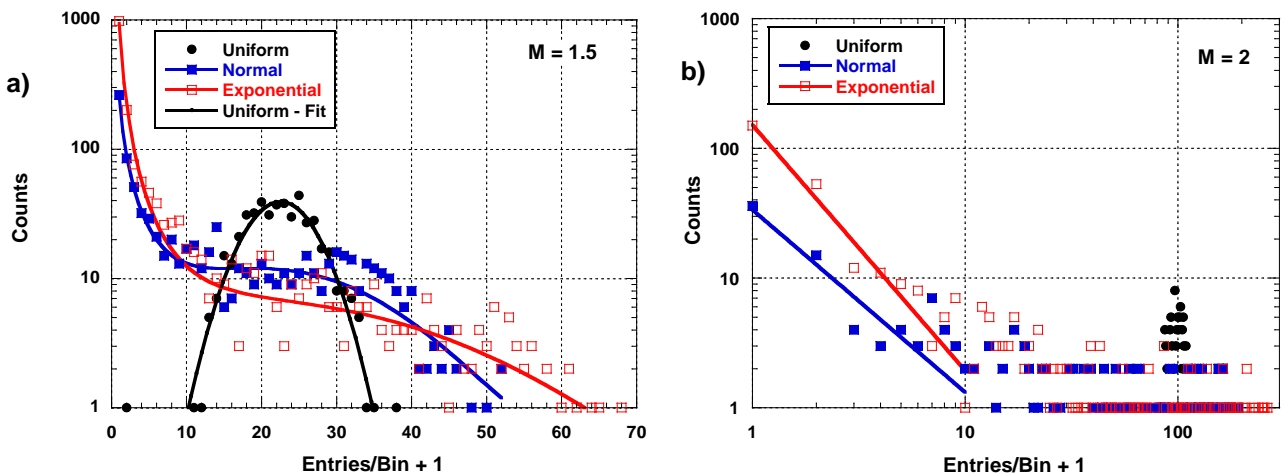

Fig. 2 Entries per bin ("EntBin") plot for uniform (closed circle), normal (closed square) and exponential (open square) distributions, generated with a total of 10,000 entries. a) M = 1.5. The points have been fitted to a normal distribution for the closed circle. The other two curves are a power law plus normal with their mean fixed to the closed circle value of 22.5. b) M = 2.0. Note the use of a log-log scale. The power law part of the plot has been fitted up to entries/bin of 9. The fitted parameters for both plots are given in Table. 2.1. Note that the maximum entries/bin is roughly twice the peak position of the uniform distribution, for the normal and exponential distributions. See text for a full discussion.

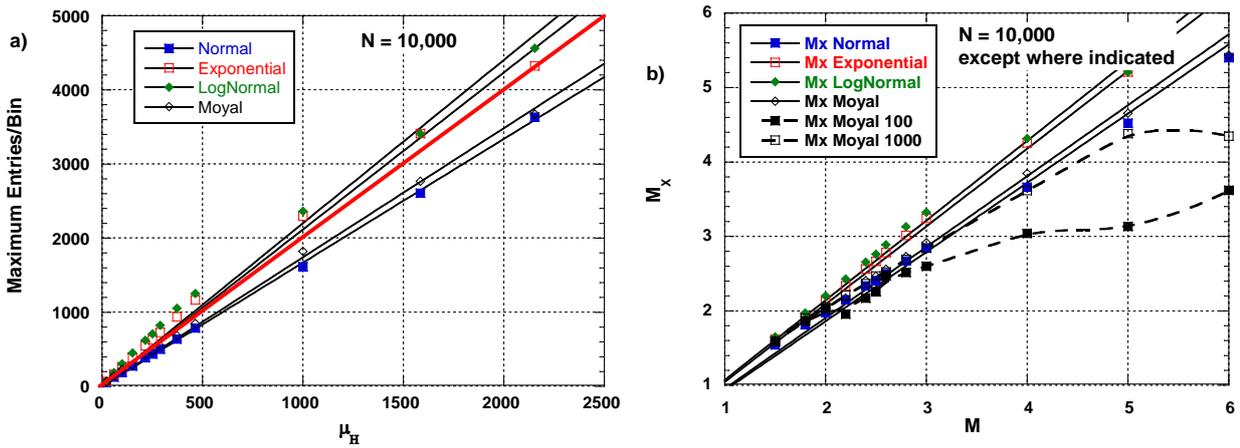

Fig. 3 a) Maximum entries per bin in histogram versus the mean number of entries per bin defined by the parameter M in the histogram entropy, $H = (1/M)\log N$. M varies from 1.5 to 6. The lines have slopes, 1.67 +/- 0.01, 2.12 +/- 0.04, 2.20 +/- 0.05 and 1.74 +/- 0.02. These are for normal, exponential, log-normal and Moyal distributions respectively. The central line is for a slope of two. b) $M_X$ estimated using Eq. (24) versus actual M for same data as in a). The Moyal distribution data is plotted for N = 100, 1000 and 10,000. See text for full discussion.

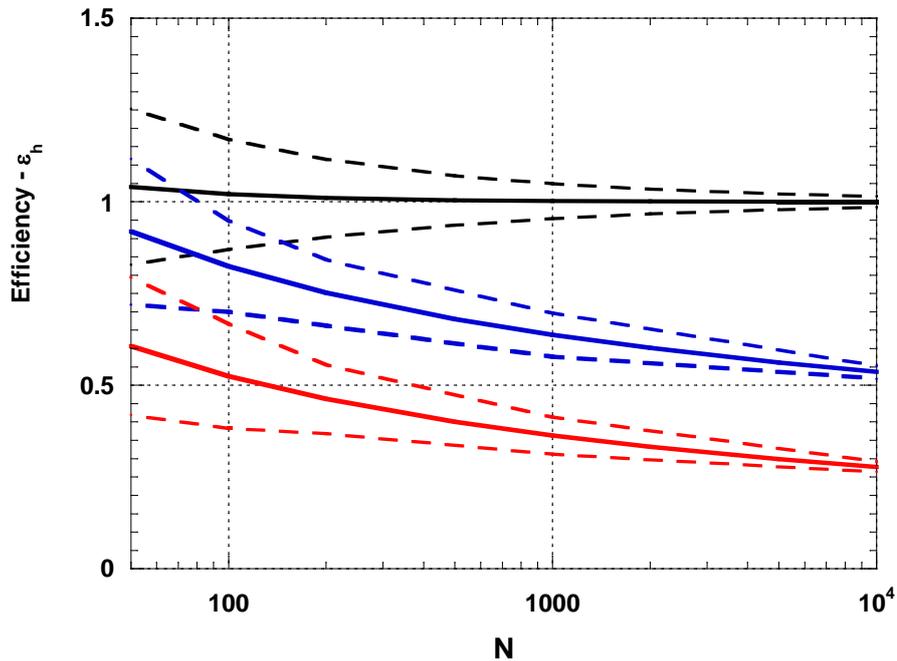

Fig. 4 Efficiency, $\varepsilon_h$, (bold lines) versus sample size, for uniform (top), normal (middle), and exponential (bottom) distributions, using Eq. (30). These curves apply to any uniform, normal or exponential distribution as the efficiency is scale independent. One standard deviation (broken lines) limits are shown. For the uniform distribution, the error on the efficiency is the same as the error on h ( nats). The error on h is approximately the same for any distribution. The error on the efficiency for other distributions is also affected by the error on the range. See text for detail.

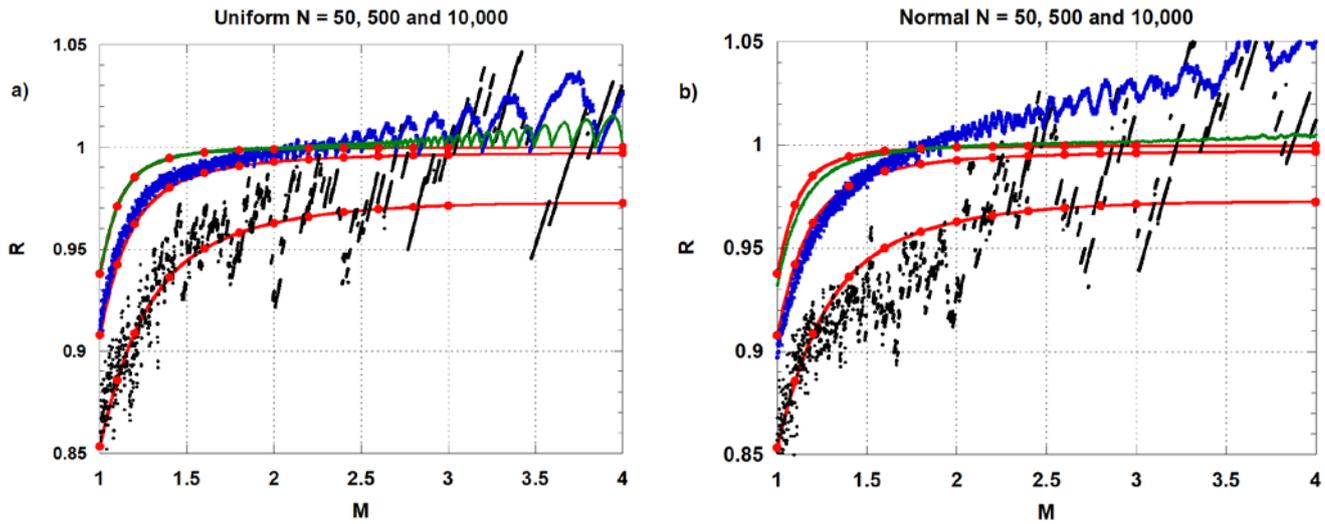

Fig. 5 Ratio, R, of the binned entropy to input entropy set by M, for uniform, a) and normal distribution b). Drawn for different number of samples – 50, 500 and 10,000. The curves are the theoretical prediction, Eq. (33) with calculated points marked with a filled circle. N = 50, 500 and 10,000 curves and simulated data start at R = 0.85, 0.91 and 0.94 for M = 1.

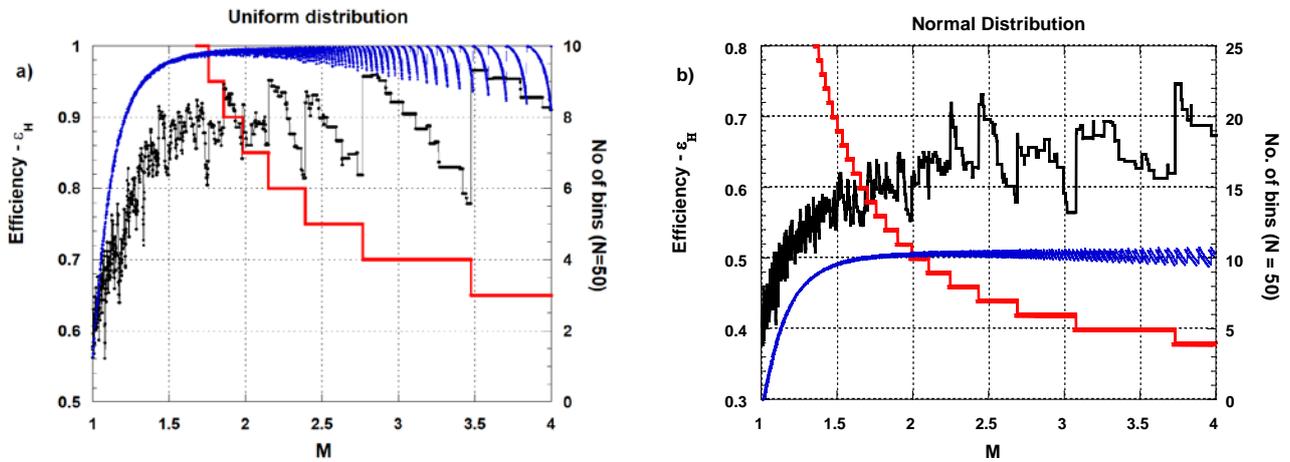

Fig. 6 Efficiency, $\varepsilon_H$, for uniform, a) and normal, b) distributions. Plotted for N = 50 and 10000. The number of bins for N = 50 is shown on the right y-axis. It is a decreasing function of M. The efficiency increases with M and saturates for M > 2. The fluctuations in efficiency are smaller for larger N. See text for discussion.

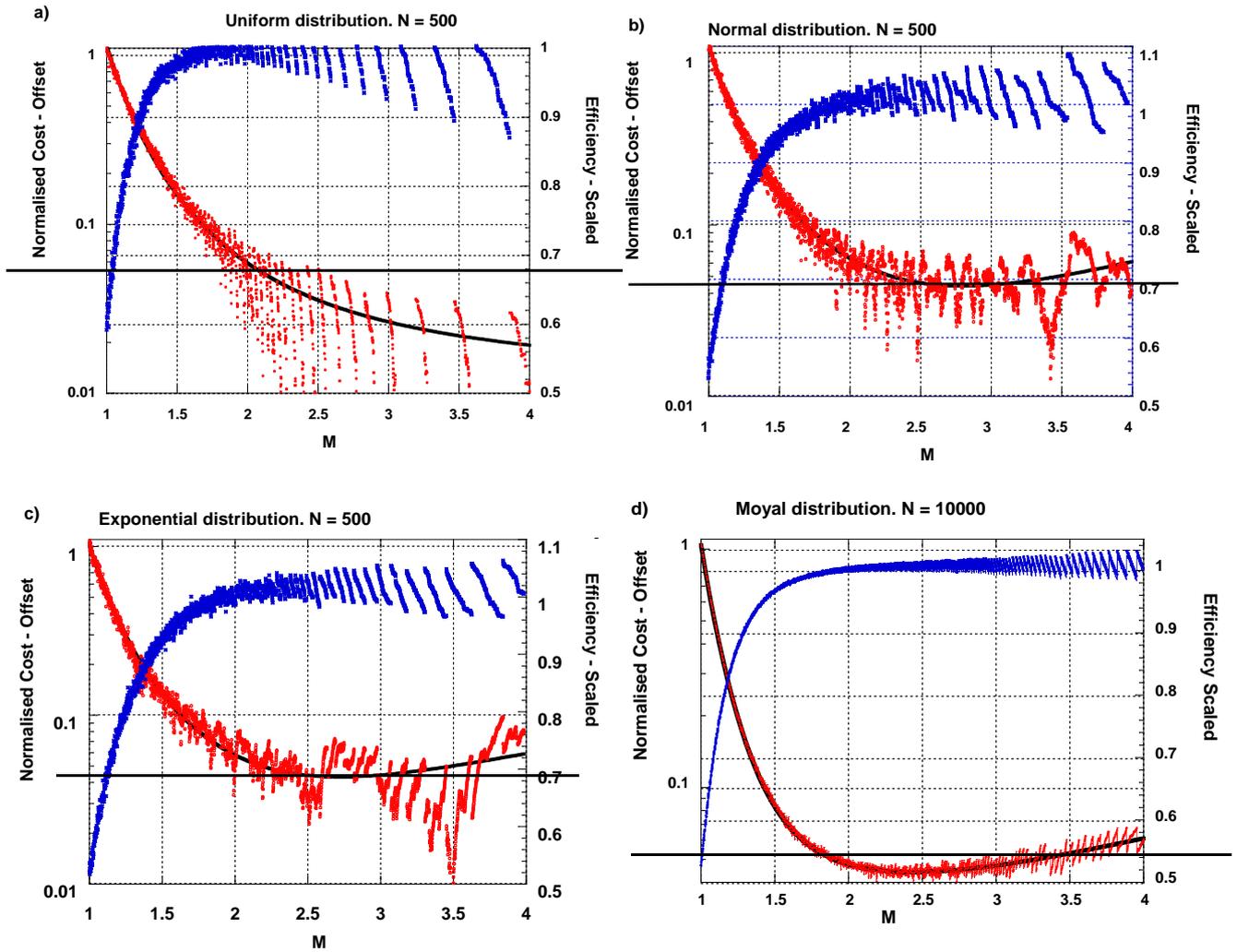

Fig. 7 Normalized cost, offset by 0.05, and efficiency plots for uniform a), normal b), and c) exponential distributions for N = 500. The Moyal distribution is shown in d) for N = 10,000, to compare with the histograms in Fig. 1. Offset of 0.05 shown by line to indicate actual zero on the log-scale. Efficiency, $\varepsilon_H$, prior to scaling is, 0.97, 0.63, 0.47 and 0.43 for a) to d) respectively. The curves are a fit to the cost using Eq. (38).

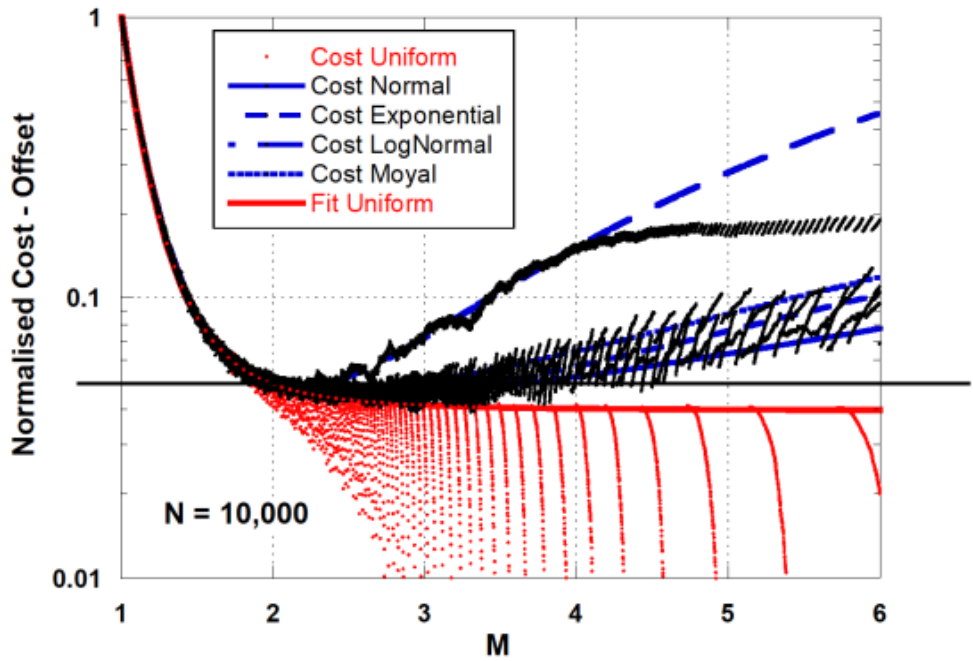

Fig. 8 Normalised cost versus M. This is offset by 0.05 indicated by the horizontal line. From top to bottom on the right hand side axis the distributions are, log-normal, Moyal, exponential, normal and uniform. Fit parameters are given in Table 4.1.

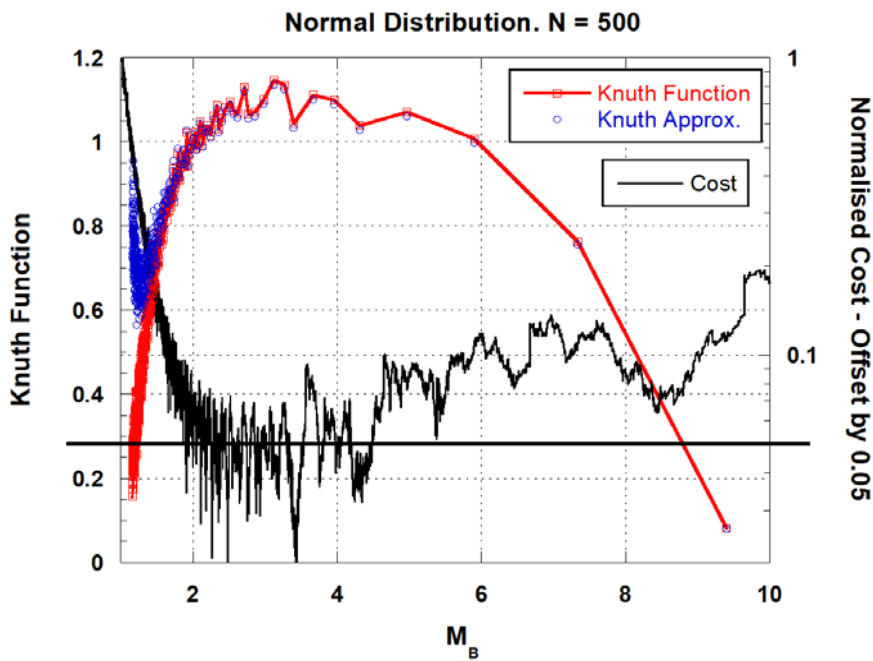

Fig. 9. For normal distribution with N = 500. Top curve - Knuth posterior probability function, Eq. (44) versus $M_B$ derived from Eq. (23). Approximate function, Eq. (45), open points. Bottom curve is the normalised cost function offset by 0.05. The offset is shown by the horizontal line to indicate the actual zero.

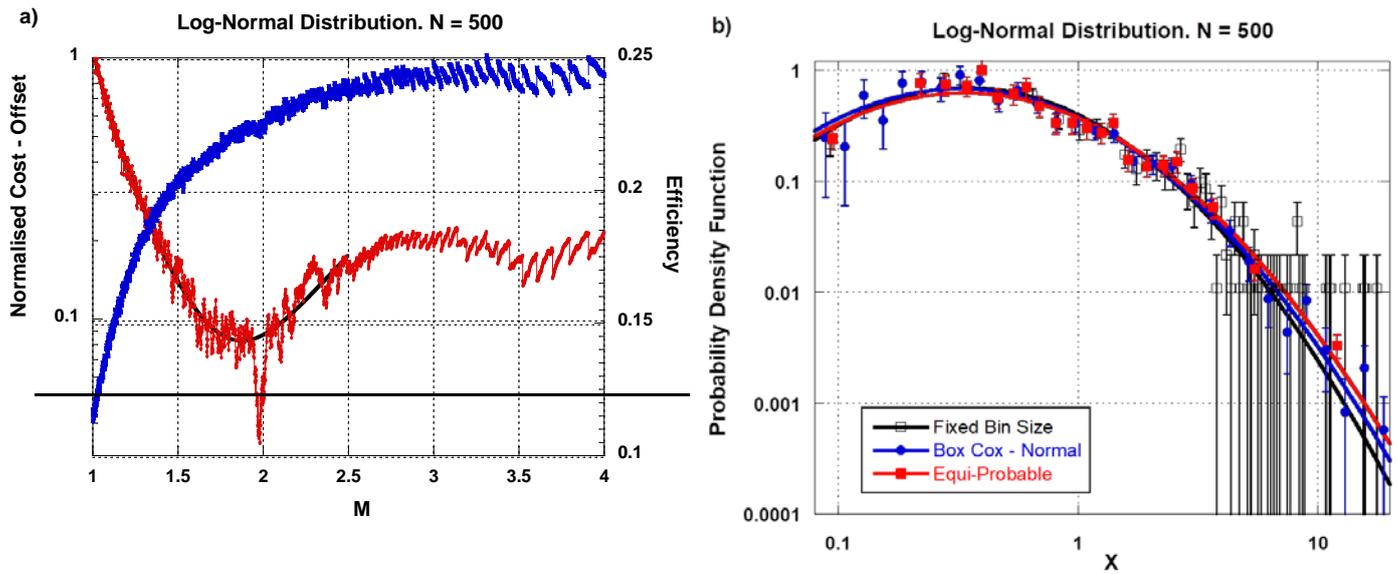

Fig. 10 a) Normalised cost function and efficiency, $\varepsilon_H$ for the log-normal distribution, with N = 500. b) Probability density function for the log-normal data using fixed binning ( open square), Box-Cox transform to normal ( closed circle) and equiprobable binning ( closed square). M = 2 for all types of binning.

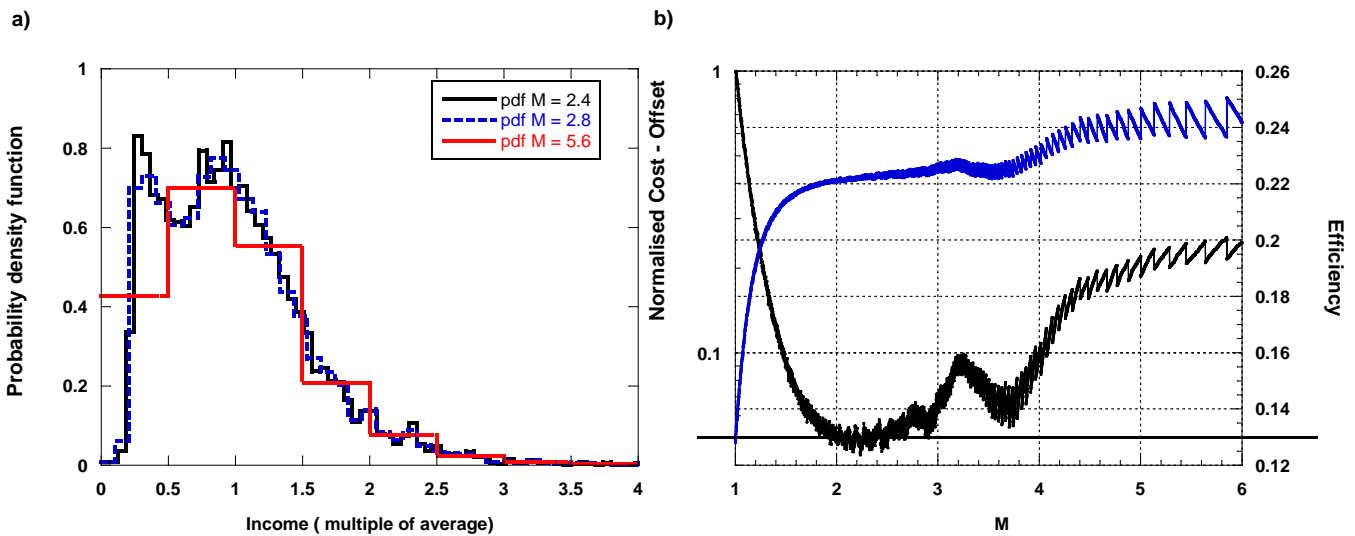

Fig. 11 a) Histogram of the income data for different values of M that match the bin widths used by Wand [7]. M = 2.4 corresponds to the algorithm used by Wand which improves on Scott's rule which has M = 2.8. The M = 8 is the S-Plus default binning also shown by Wand. b) The normalised cost and efficiency for this data versus M. Offset of 0.05 shown by line to indicate zero on the log-scale.

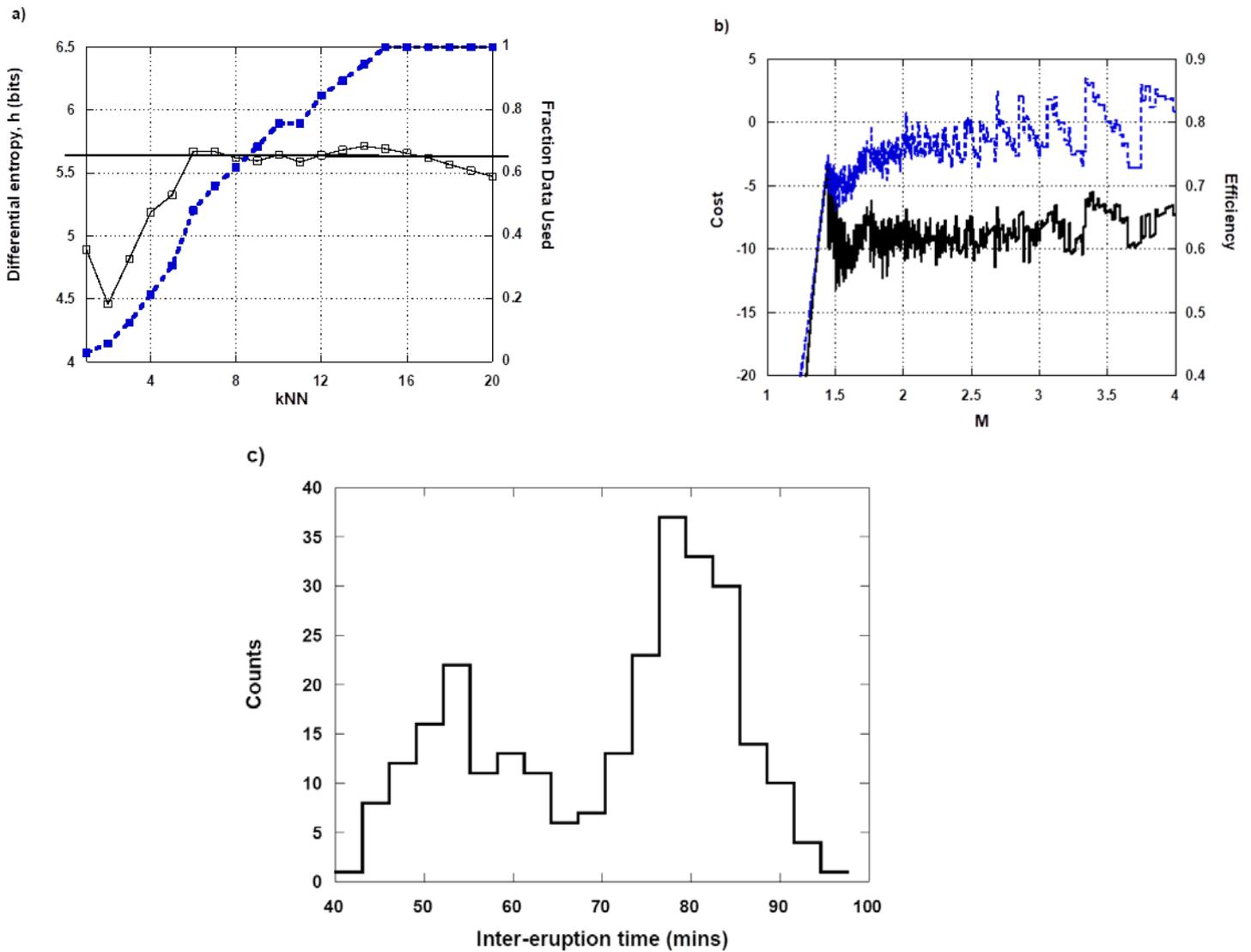

Fig. 12. Analysis of the Old Faithful geyser data, [24]. a) Estimate of differential entropy, h, for different values of nearest neighbour. The fraction of points used is on the right axis. This data is much affected by repeated values and a reliable estimate of h needs kNN > 8. The horizontal line is at 5.64 which is used for the histogram. b) Cost ( bottom line) and efficiency (top line) for this data. The lack of precision on the data leads to the discontinuity in both variables at M ~ 1.4. c) Histogram of the old faithful geyser data with M = 2. This binning correctly shows two peaks as is well known for this system.

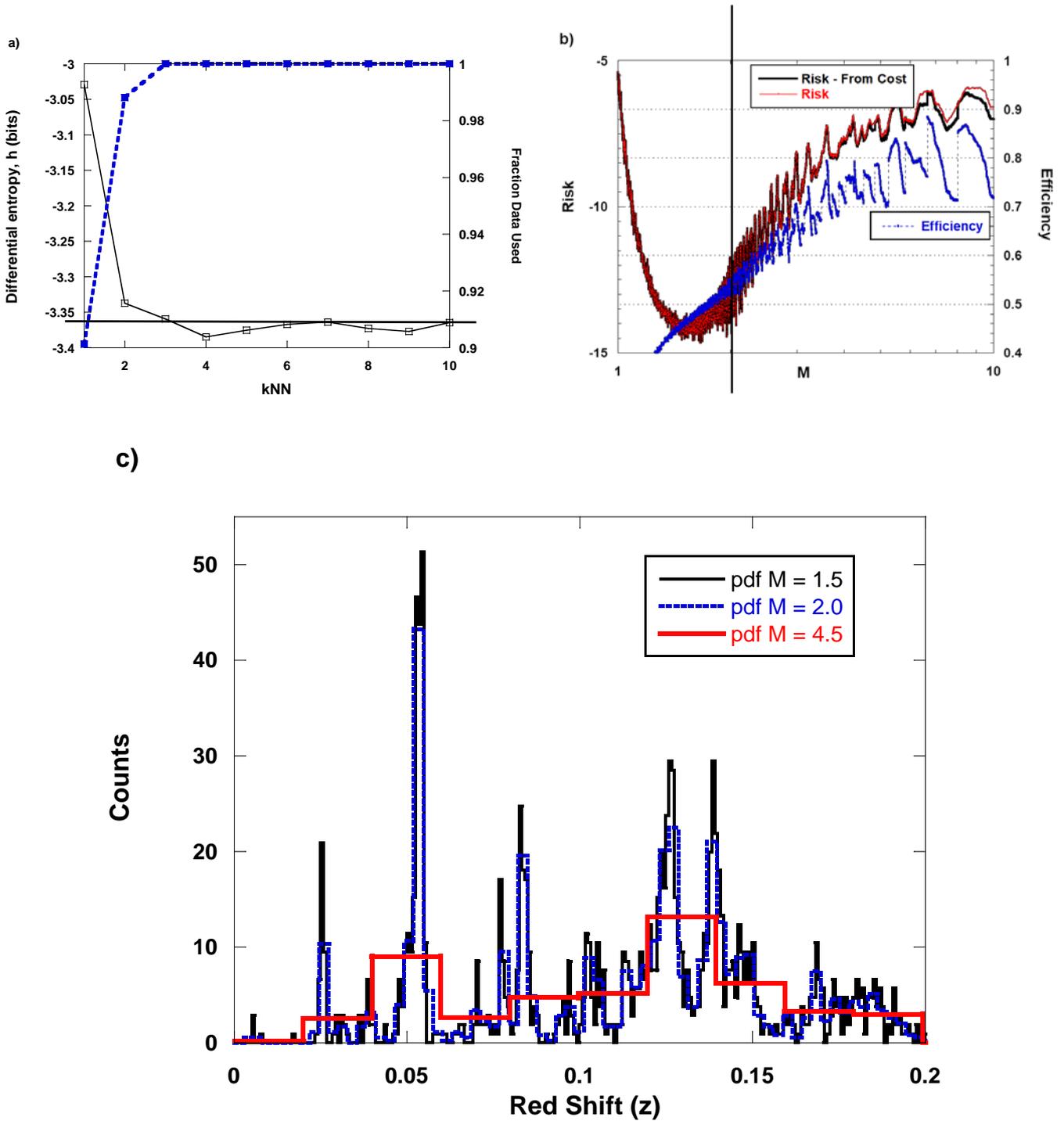

Fig. 13. Red shift of galaxies from an astronomical sky survey. Data collated by Wasserman, [26]. a) Differential entropy estimated for different values of nearest neighbour, kNN. The line at -3.36 is the one used for the histogram. b) Risk derived from the cost, Eq. (43) and the risk function of Stone and Rudemo (left axis) plus the efficiency (right axis) versus M. The x-axis has a log scale to make the data easier to study. A vertical line at M = 2 indicates the preferred value. c) Probability density function of the data for M = 1.5, 2.0 and 4.5 to be compatible with the analysis of Wasserman, [26]. The bin widths are 8.3 $10^{-4}$, 2.74 $10^{-3}$, and 1.99 $10^{-2}$ for M = 1.5, 2.0 and 4.5 respectively.

# Tables

**Table 1.1** *Histogram algorithms used by major data analysis software. Refs. [11,12,13], describe the scientific software, Numpy, R and Mathematica respectively. Two columns indicate if the algorithm uses the number of entries, N, or a feature of the pdf.*

| Method, Year and Ref. | Used by Refs. [10-13] | N | pdf | Formula (F) or Other |
|---|---|---|---|---|
| Sturges (1926) [1] | Numpy/R/Mathematica | Y | N | F – First published algorithm |
| Scott (1979) [2] | Numpy/R/Mathematica | Y | Y | F – Integrated Mean Square Error (IMSE). Formula and Rule. |
| Doane (1979) [3] | Numpy | Y | Y | F – Improvement on Sturges |
| Freeman-Diaconis (1981) [4] | Numpy/R/Mathematica | Y | Y | F - Modification of Scott's Rule. |
| Rudemo (1982) [5] | - | - | Y | Risk Function – IMSE |
| Stone (1984) [6] | Numpy | - | Y | Risk Function – IMSE |
| Wand (1997) [7] | Mathematica | Y | Y | Update of Scott - IMSE |
| Knuth (2006) [8] | Mathematica | Y | Y | Posterior Probability Function |
| Shimzaki and Shinomoto (2007) [9] | - | - | Y | Cost Function - IMSE |
| Excel [10] | Microsoft | Y | N | Number of bins is square root of N |
| This paper | | Y | Y | F |

**Table 1.2** *Differential entropy, h, bin width using Scott's Formula, and entropy for some typical pdfs.*

| Distribution | $p(x)$ | $h$ (nats) | $\Delta$ | H (nats) |
|---|---|---|---|---|
| Normal | $N(\mu,\sigma)$ | $\frac{1}{2}\log(2\pi e \sigma^2)$ | $2 \times 3^{\frac{1}{3}} \pi^{\frac{1}{6}} \sigma N^{-\frac{1}{3}}$ | $\log\left(1.1839 N^{\frac{1}{3}}\right)$ |
| Exponential | $\frac{1}{\lambda}e^{-\frac{x}{\lambda}}, \lambda > 0$ | $1 + \log(\lambda)$ | $12^{\frac{1}{3}} \lambda N^{-\frac{1}{3}}$ | $\log\left(1.1873 N^{\frac{1}{3}}\right)$ |
| Maxwell-Boltzmann | $4\pi^{-\frac{1}{2}} \beta^{\frac{3}{2}} x^2 e^{-\beta x^2}$ $x, \beta > 0$ | $\frac{1}{2}\log\left(\frac{\pi}{\beta}\right) + \gamma - \frac{1}{2}$ $\gamma = 0.5772$ Eulers Constant | $1.6258 \beta^{-\frac{1}{2}} N^{-\frac{1}{3}}$ | $\log\left(1.1777 N^{\frac{1}{3}}\right)$ |
| Triangular | $\frac{2x}{a} \quad 0 \le x \le a$ $\frac{2(1-x)}{1-a} \quad a \le x \le 1$ | $\frac{1}{2} - \log(2)$ | $\left\{\frac{6a(1-a)}{4}\right\}^{\frac{1}{3}} N^{-\frac{1}{3}}$ $0 < a < 1$ | $\log\left(1.607 N^{\frac{1}{3}}\right),$ $a = 0.1$ $\log\left(1.1432 N^{\frac{1}{3}}\right),$ $a = 0.5$ |
| Uniform | $\frac{1}{\beta}, 0 \le x \le \beta$ | $\log(\beta)$ | Infinite | See text |

**Table 2.1** *Fit parameters to Fig. 2. See text for detail.*

| Fit to Fig. 2a M = 1.5 | Power Law + Normal Distribution - $N_{Empty}(X+1)^{-\alpha} + S \times N(\mu,\sigma)$ | | | | | |
|---|---|---|---|---|---|---|
| pdf | $N_{Empty}$ | $\alpha$ | S | $\mu$ | $\sigma$ | $\chi^2$/DF |
| Uniform | - | - | 39.5+/-2.3 | 22.5+/-0.2 | 4.5+/-0.2 | 0.8 |
| Normal | 258.9+/-15 | 1.56+/-0.08 | 9.8+/-0.9 | 22.5 Fixed | 12.7+/-1.0 | 1.15 |
| Exponential | 941+/-30 | 2.1+/-0.05 | 5.4+/-0.6 | 22.5 Fixed | 21.0+/-2.0 | 1.48 |
| Fit to Fig. 2b M = 2.0 | Only Power Law distribution at low X | | | | | |
| Normal | 33.6+/-5.6 | 1.40+/-0.2 | - | - | - | 1.2 |
| Exponential | 151.7+/-4.9 | 1.89+/-0.1 | - | - | - | 1.65 |
| Fit to Fig. 2b M = 2.0 | Only Normal distribution at high X | | | | | |
| Uniform | - | - | 4.5 +/-0.4 | 100.0+/-0.8 | 8.54+/-0.8 | - |

**Table 2.2** *Check on Eq. (29) to calculate $K_{Norm}$ from which $\alpha_{Pred}$ is estimated using Eq. (27). $\alpha_{Actual}$ from Table 2.1. Empty bin count directly from histogram. Efficiency estimated from the histogram. N = 10,000. Shaded columns calculated from those to left.*

| Fig. 2b M = 2.0 Check of Eq. (29) | $N_{Empty} = \left(\dfrac{1-\varepsilon}{\varepsilon}\right) N^{\frac{1}{M}} \times \dfrac{1}{K_{Norm}}$ | | | | | |
|---|---|---|---|---|---|---|
| pdf | $\varepsilon$ | $(1-\varepsilon)/\varepsilon$ | $N_{Empty}$ | $K_{Norm}$ | $\alpha_{Pred}$ | $\alpha_{Actual}$ |
| Normal | 0.506 | 0.978 | 36 | 2.72 | 1.26 | 1.40 |
| Exponential | 0.260 | 2.840 | 150 | 1.74 | 1.83 | 1.89 |

**Table 4.1** *Fit parameters to the cost data shown in Fig. 8. The fit function is Eq. (37). See text for detail.*

| Fig. 8 Fit to Eq. (37) | $C_{NR} = m_1 N^{\frac{1}{M}-1} - m_2 N^{-\frac{1}{2}} + m_3(M-|m_4|)^2 + 0.05$ | | | | |
|---|---|---|---|---|---|
| Parameter | Uniform | Normal | Exponential | Log-normal | Moyal |
| a) Fit between M = 1 and M = 2 only | | | | | |
| $m_1$ | 0.998+/-0.0006 | 1.028+/-0.0006 | 1.014+/-0.0006 | 1.01+/-0.0008 | 1.02+/-0.0006 |
| $m_2$ | 1.07+/-0.01 | 1.1 +/- 0.01 | 1.16 +/-0.01 | 0.981+/-0.02 | 0.9335+/-0.01 |
| Regression R | 0.99983 | 0.99985 | 0.99983 | 0.99967 | 0.99985 |
| b) Fit between M = 1 and M = 4. All parameters. Fit to uniform not needed. | | | | | |
| $m_1$ | - | 1.03+/-0.0005 | 1.01+/-0.0007 | 0.985+/-0.0008 | 1.02+/-0.0007 |
| $m_2$ | - | 1.16+/-0.015 | 1.15+/-0.01 | 1.13+/-0.01 | 0.973+/-0.01 |
| $m_3$ | - | (1.76+/-0.1)x10$^{-3}$ | (3.7+/-0.1)x10$^{-3}$ | (24.4+/-0.1)x10$^{-3}$ | (4+/-0.1)x10$^{-3}$ |
| $m_4$ | - | 1.28+/-0.07 | 1.81+/-0.03 | 1.86+/-0.01 | 1.58+/-0.03 |
| Regression R | - | 0.99982 | 0.99964 | 0.99957 | 0.99972 |

# Appendix

## A1. Probability distribution functions simulated for this work

The following table shows the key characteristics of the distributions simulated for the work in this paper.

| Distribution | Mean | Standard Deviation | Support | Skew | Kurtosis | h (bits) | Generator Name/Method |
|---|---|---|---|---|---|---|---|
| Standard Normal | 0.0 | 1.0 | R | 0.0 | 3.0 | 2.047 | G<br>Box-Muller |
| Uniform | 0.5 | 0.3 | (0,1) | 0.0 | -1.2 | 0.0 | U<br>See text |
| Exponential | 1.0 | 1.0 | (0,Infinity) | 2.0 | 6.0 | 1.45 | ---<br>$-\lambda \ln(U)$ |
| Standard Log-Normal | 1.65 | 2.16 | (0,Infinity) | 6.18 | 111 | 2.047 | ---<br>$\exp(G)$ |
| Moyal | 1.28 | 2.21 | R | 1.4 | 3.0 | 2.96 | ---<br>$-\log(G^2)$ |

The pdf for three of these distributions are given in Table 1.2 in the main text. The Moyal distribution is an approximation to the Landau distribution which describes the energy loss fluctuations of a charged particle in a medium. The Moyal distribution from ref. [A1] is simple to generate. The pdf is given by,

$$p(x) = \frac{1}{\sqrt{2\pi}} \exp(-\frac{1}{2}(x + \exp(-x)))$$

This is effectively a normal distribution with a right hand exponential tail. It has no free parameters.

The standard Log-Normal has a pdf given by,

$$p(x) = \frac{1}{x\sqrt{2\pi}} \exp(-\frac{(\log x)^2}{2})$$

Generation of all distributions starts with a uniform random number, U, between zero and one downloaded from ref. [A2] which uses atmospheric noise. A normal or Gaussian number, G, can be generated using the Box-Muller method, [A3]. The other distributions are generated from either U or G as indicated in the table. These formula are derived by inversion from the cumulative density function.

Plots and the results of curve fitting came from the use of KaleidaGraph, ref. [A4].

## A2. Normalised cost function

$$C_{SS} = \frac{2\mu_B - \sigma_B^2}{\Delta^2}$$

In the region where Poisson fluctuations dominate ( $1 \leq M \leq 2$ )

$\sigma_B^2 \sim \mu_B$ thus $C_{SS} \approx \frac{\mu_B}{\Delta^2}$

Using the equations in the main paper, repeated below,

$\mu_B = \varepsilon \mu_H$  $\mu_H = N^{1-\frac{1}{M}}$ and $\Delta = \frac{2^h}{N^{\frac{1}{M}}}$

So $C_{SS} = f(h)\varepsilon \mu_H N^{\frac{2}{M}}$ where $f(h) = 2^{-2h}$

The maximum value of $C_{SS}$ is at M = 1. Fluctuations will be gone by M = 2. Taking reference points for $C_{SS}$ at M = 1, $C_1$, and $C_{SS}$ at M = 2, $C_2$, define a normalised cost function, with the efficiency at these points given by $\varepsilon_1$ and $\varepsilon_2$, and note that $\mu_H$ is one and $N^{\frac{1}{2}}$ at M = 1 and M = 2 respectively.

$C_N = \frac{C_{SS} - C_2}{C_1 - C_2} = \frac{\varepsilon \mu_H N^{\frac{2}{M}} - \varepsilon_2 N^{\frac{1}{2}} N^1}{\varepsilon \mu_H N^{\frac{2}{M}} - \varepsilon_2 N^{\frac{1}{2}} N^1} = \frac{\varepsilon \mu_H N^{2(\frac{1}{M}-1)} - \varepsilon_2 N^{-\frac{1}{2}}}{\varepsilon_1 - \varepsilon_2 N^{-\frac{1}{2}}}$

$\varepsilon_2 N^{-\frac{1}{2}}$ is small compared to the other term in the denominator. The only term with a pdf dependence f(h) cancels out in the normalised cost function.

$C_N = \frac{\varepsilon}{\varepsilon_1} \frac{1}{\mu_H} - \frac{\varepsilon_2}{\varepsilon 1} N^{-\frac{1}{2}} = m_1 \frac{1}{\mu_H} - m_2 N^{-\frac{1}{2}}$

Expect m₁ and m₂ to be approximately one with m₂ slightly larger than m₁ .

**A3. Relationship between Shimazaki and Shinomato work and Scott's Formula**

Shimazaki and Shinomato find two asymptotic solutions to the cost function, using the autocorrelation of the pdf,

$\varphi(\tau) = \int_S p(x) p(x-\tau) dx$ (E1)

Using a Taylor's Series expansion, one can write,

$p(x-\tau) = p(x) - \tau p'(x) + \frac{\tau^2}{2} p''(x) + \ldots$

Using this expansion in Eq. (E1) one obtains,

$\varphi''(\tau) \sim \int_S p(x) p''(x) dx$ (E2)

Using the integration by parts formula, $\int uv' dx = uv - \int vu' dx$ and set $u = p(x)$ and $v = p'(x)$,

$\int_S p(x) p''(x) dx = \left[ p(x) p'(x) \right]_{x_{Low}}^{x_{High}} - \int_S p'(x) p'(x) dx$

For pdfs with p(x) = zero at $x_{Low}$ and $x_{High}$, and a reasonably behaved p(x), ensures a symmetric autocorrelation behaviour at τ = 0,

$\varphi''(\tau) \sim \int_S p(x) p''(x) dx = -\int_S p'(x)^2 dx$

Thus, Shimazaki and Shinomoto's formula for a symmetric autocorrelation at τ = 0 can be written,

$$\Delta \sim \left[ -\frac{6}{\phi''(0)N} \right]^{\frac{1}{3}} = \left[ \frac{6}{\left\{ \int p'(x)^2 dx \right\} N} \right]^{\frac{1}{3}}$$ which is the same as Scott's Formula.

**Appendix References**

[A1] J.E. Moyal (1955) XXX. *Theory of ionization fluctuations*, The London, Edinburgh, and Dublin Philosophical Magazine and Journal of Science, 46:374, 263-280, https://doi.org/10.1080/14786440308521076

[A2] Haahr, M. (2022). *RANDOM.ORG: True Random Number Service*. https://www.random.org/ Last accessed 26 July 2022.

[A3] Box, G. E. P.; Muller, Mervin E. (1958). *A Note on the Generation of Random Normal Deviates*. The Annals of Mathematical Statistics. 29 (2): 610–611. https://doi.org/10.1214/aoms/1177706645

[A4] KaleidaGraph, Version 4.5.4 for Windows. Synergy Software, Reading, PA, USA. www.synergy.com.